# Spontaneous formation of well-defined Al rich shell structures in $Al_xGa_{1-x}N$/GaN nanowires

*D. González[1] R. Fath[1] T. Ben[1], and R. Songmuang[2]*

[1] Departamento de Ciencia de los Materiales e I. M. y Q. I., Universidad de Cádiz, Puerto Real, Cádiz, Spain.

[2] CEA-CNRS Group Nanophysics and Semiconductors, Institute Néel, 17 Rue des Martyrs, 38054 Grenoble cedex 9, France

E-mail: david.gonzalez@uca.es

Growth of catalyst-free $Al_xGa_{1-x}N$ ($0.15<x<0.50$)/GaN nanowires by plasma assisted molecular beam epitaxy is thoroughly structural and chemical analyzed by using transmission electron microscopy related techniques. We found that well-defined and defect-free core-shell structures are spontaneously formed during the wire growth. An Al-rich shell with significantly higher Al composition pseudomorphically encapsulates a Ga-rich $Al_xGa_{1-x}N$ core with an atomically-abrupt hetero-interface. Nevertheless, the energy dispersive X-Ray spectroscopy reveals a complex chemical composition gradient along the wire axis for both core and shell blocks which is ascribed to the adatom surface kinetic differences and the shadow effect during the growth

Group III-nitride semiconductors are the materials of choice for large varieties of optoelectronic applications, especially light emitting/detecting devices in green to ultraviolet spectral region. Generally, conventional thin films of this material family are grown on non-native substrates, leading to a high density of defects which deteriorates the device performance. Lately, nanowires have emerged as an alternative solution to suppress defect formation in III-nitrides because of the efficient elastic strain management in this one-dimensional geometry in spite of the large lattice mismatch growth [1].

Furthermore, the heterostructures inside III-N nanowires is also a promising building block for the quantum dot fabrication [2,3,4]. However, the built-in electric field induced by the polarization difference in GaN/AlN heterostructures grown along [0001] direction could give an undesirable effect; for example, it could drastically reduce the luminescence efficiency of the quantum dots due to quantum confined stark effect [5]. Using ternary alloy heterostructures in nanowires such as GaN/AlGaN [6] or AlGaN/AlN would thus give an opportunity not only to tailor the bandgap of the material but also to attenuate the internal electric field in the structures. Despite their necessity, so far, there is only a few attempts to understand the growth of ternary alloy nanowires in order to precisely design and control their structures and chemical compositions [7,8,9].

In this work, the growth of catalyst-free $Al_xGa_{1-x}N$/GaN nanowires by plasma assisted molecular beam epitaxy (PAMBE) was studied. Their structural quality and composition distribution in nanometer scale were analyzed by using various transmission electron microscopy techniques. Our investigations reveal that Al-rich core-shell like structures spontaneously form during the growth of $Al_xGa_{1-x}N$ sections in the nanowires. The exact composition and dimension of both core and shell along the structure is carefully measured. Moreover, the measured thicknesses and chemical compositions were compared with the values directly derived from the growth conditions. Finally, we propose a simple model to describe the situation occurring during the growth of $Al_xGa_{1-x}N$ in the nanowire.

Our starting point is catalyst-free GaN nanowires grown by using PAMBE on Si(111) substrates under N-rich atmosphere[10]. $Al_xGa_{1-x}N$ sections with the length of 200-300 nm with a single nominal Al composition ($x_{Al}$) were grown on the top of those GaN nanowires. High-resolution transmission electron microscopy (HRTEM) was performed using a JEOL 2011 microscope. Compositional analysis was performed using a JEOL JEM-2010F Scanning Transmission Electron Microscopy (STEM), which was equipped with Energy Dispersive X-ray spectroscopy (EDX) and Z-contrast High Angle Annular Dark Field (HAADF) modules.

In order to estimate $x_{Al}$ in $Al_xGa_{1-x}N$ section in nanowires, we apply an approximation concept similar to that used in the two dimensional layer systems. In that case, knowing that the binding energy of Al-N bond is larger than that of Ga-N bond, $x_{Al}$ is then equal to $R_{Al}/R_N$ if $R_{Al}+R_{Ga}>R_N$, or in the other word, if the deposition was done under metal-rich conditions[11]. $R_{Al}$ and $R_{Ga}$ are the impinging Al and Ga fluxes while $R_N$ is the supplied nitrogen flux (in our case ~0.8 Å/s). In our experiment, the GaN nanowires were intentionally grown in the regime where the nanowire axial growth rate is limited by the supplied nitrogen flux, implying the local Ga rich condition on the wire top[12]. Therefore, $x_{Al}$ in $Al_xGa_{1-x}N$ in nanowires can be estimated by $R_{Al}/R_N$ as in the thin film. Note that $R_{Al}$ in nanowires is, in fact, equivalent to that of two-dimensional layer since the diffusion of Al adatom along nanowire sidewall is negligible at our growth temperature[9]. Moreover, the thickness of GaN and $Al_xGa_{1-x}N$ sections in nanowires can be deduced by $R_N \cdot t$, where $t$ is the deposition time. We will show later on that this assumption can provide an average value of Al content without considering the alloy fluctuation as well as insertion thicknesses.

Figure 1 shows an HAADF STEM image of a single $Al_xGa_{1-x}N$/GaN nanowire with the estimated $x_{Al}$=0.25. The GaN base is shown as the brightest contrast. For $Al_xGa_{1-x}N$ section, a clear contrast of a bright inner core compared to a dark outer shell is observed and this behavior appears in all of the investigated nanowires with different $x_{Al}$ ranging from 0.15 to 0.53. This dark shell layer, presumably richer of Al atoms, entirely wraps around the poorer Al core. In fact, the shell-like structure was also observed continuously on the top of the GaN nanowire basement. Typically, the measured cores have diameters in the range of 20-30 nm, whereas the shell thicknesses are about 10 nm near the GaN/Al$_x$Ga$_{1-x}$N interface. The HRTEM image taken along the [11-20] pole at the edge of nanowire indicated by the square (shown by the inset of Fig. 1) unmistakably evidences a sharp interface separating the core and shell regions along the nanowire. In this micrograph, the shell follows the same crystal structure of the core with an abrupt contrast change at the interface. Importantly, the shell exhibits a pseudomorphic relationship regarding the core at the interface all along the whole nanowire. There is no evidence of disorientations or any other crystalline defects such as grain boundaries or misfit dislocations.

Figure 2 shows the average thickness profiles of the shell ($d_{shell}$), the core ($D_{core}$) and the whole nanowire diameter ($D_{tot}$) along the Al$_x$Ga$_{1-x}$N/GaN nanowire growth direction. The well-defined interface between the core and the shell assists the analysis of their thicknesses with a nanometer precision. In most of the investigated wires, the shell thickness begins at around 300 nm beneath the GaN/Al$_x$Ga$_{1-x}$N interface, then slowly increases to the maximum value of about 10 nm and finally decays to zero close to the wire apex. There is no evidence of a shell covering the GaN in the region near to the substrate. This behavior is attributed to the shadow effect occurring when the nanowire density is high. Differently, $D_{core}$ shows a tendency to increase up to the apex, where a separation between the rich and poor Al phases cannot be distinguished. In any cases, the shell thickness participates with a high percentage in the whole nanowire diameter.

Considering that only two atoms such as Al and Ga are brought to meet stoichiometry with nitrogen, we have analyzed the Ga/Al ratio by using spot EDX measurements along the growth direction on several nanowires. Since there is a significant difference in the composition between the core and shell regions, it is necessary to separate both contributions to depict a clear image of the Al distribution in the nanowires. For this, EDX spot measurements were made separately at the same height to analyze both Al content in the shell regions ($x_{shell}$) and in the nanowire centers ($x_{center}$). Additionally, the thickness measurements of the core and shell were carried out simultaneously[13]. We assume that a gradient of Al composition along the diameter does not occur, since the EDX profiles across the diameter follow a typical thickness-dependence. Thus, we have estimated the Al composition inside the core ($x_{core}$) by

$$x_{core} = \frac{x_{center} \cdot D_{tot} - x_{shell} \cdot 2d_{shell}}{D_{core}} \quad (1)$$

The $x_{shell}$, $x_{center}$ and $x_{core}$ along the nanowire axis are plotted in Fig. 3(a). The measured $x_{shell}$ starts from the expected value of 0.25, and then it increases to 0.82 at the area close to the GaN/Al$_x$Ga$_{1-x}$N interfaces. This plot highlights the extremely high Al content of the shell regarding to the center value. In the Al$_x$Ga$_{1-x}$N area, the Al content in the shell remains almost constant and it gradually decreases at the nanowire top region. This feature appears at the same point where the shell thickness starts to decrease. Differently, the measured $x_{center}$ oscillates in the range of 0.28-0.45, firstly decreasing slowly and finally undergoing a rise when approaching to the apex. From *eq. 1*, the estimated $x_{core}$ is varied in the range of 0.21-0.39 along the axis, following the same tendency as $x_{center}$. The $x_{core}$ value nearly zero in GaN nanowire basement area confirms that our approximation method is reasonable. Important to note that a similar behavior of shell compositions is measured in other samples with different nominal Al content. However, the shell does not vanish in the apex if $Al_xGa_{1-x}N$ section is shorter (around 200 nm).

Our structural analysis in $Al_xGa_{1-x}N$ section reveals the Al atom distribution in the core and the shell area differing from the designed value. However, if the mass conservation was assumed, the average value obtained by EDX measurement should be similar to the Al content of the deposited material. The validation of this concept is confirmed in Fig. 3(b) where the average $x_{center}$ along the $Al_xGa_{1-x}N$ section agrees well with the predicted value derived from the growth conditions as described before. Furthermore, the predicted thicknesses are also consistent with the measured value as shown by Fig. 3(c).

From our experimental results, we propose, in Fig. 4, schematic illustrations which represent the situation occurring during the overgrowth of $Al_xGa_{1-x}N$ section on the 500-nm GaN nanowires. It is well-accepted that both of the direct impingement on the wire top and the adatom diffusion along the wire sidewall contribute to the nanowire growth.[14,15,16] The direct deposition, in fact, occurs not only on the wire top but also at the wire sidewall, leading to the formation of $Al_xGa_{1-x}N$ shell around GaN nanowire basement [Fig. 4(a)]. Nonetheless, the growth directionality of MBE makes the latter one much smaller, resulting in the lower lateral growth rate than the vertical one. This has been also observed in the growth of GaN/AlN nanowire heterostructures.[17,18]

In the case of ternary alloy, a surface migration difference of the deposited species could cause a significant chemical composition gradient in the structure. It is known that Al atom has a higher energy barrier for a surface migration process, giving a diffusion length that is much lower than that of Ga[12,19,20]. During the direct sidewall deposition of the metallic atoms, Ga atoms at the region close to the nanowire apex have a high potential to reach and contribute to the vertical growth at the wire top, leaving a larger amount of Al atoms to incorporate at the sidewalls. This fact could lead to the formation of Al-rich shell area wrapping around Ga-rich AlGaN core and the top part of GaN nanowire basement. At the area far below the $Al_xGa_{1-x}N$/GaN interface, the decreasing $x_{shell}$ implies the low probability of Ga atoms to reach and incorporate at the top region of the wire.

During the continuous growth of $Al_xGa_{1-x}N$ section, the efficient sidewall incorporation of Al atoms progressively increases the nanowire diameters. Such a structural evolution could possibly enhance the shadow effect, implying that the deposition at the wire sidewall becomes less important as the wires are getting longer. This fact automatically reduces the amount of diffusing Ga atoms from the wire sidewall, consequently, sharpens the shell and leads to the increasing of Al content in the long wires at the end [Fig. 4(b)]. At the extreme case when the wires start to merge, the sidewall diffusion is negligible and thus the Al composition in $Al_xGa_{1-x}N$ section would be similar to the .$Al_xGa_{1-x}N$ two-dimensional layer.

In conclusion, $Al_xGa_{1-x}N$ section with $0.15<x<0.5$ grown on GaN nanowires by PAMBE exhibits a spontaneous formation of Al shell-like structure. The shell appears not only along the $Al_xGa_{1-x}N$ sections but also on the top area of the GaN nanowire basement. The complex Al composition distribution is mainly attributed to the low mobility of Al atoms in comparison to Ga atoms. The spontaneous formation of core-shell $Al_xGa_{1-x}N$ nanowires might be interesting to create self-forming core shell structure in nanowires. However, it needs to overcome two main problems: the gradually decay of the core composition along the growth direction and the total vanishing of the shell caused by the shadow effect. The exact comprehension of the mechanism involved in the formation and annihilation of the core-shell structures will allow us to design suitable growth routes to obtain axial wires with both thickness and composition control.

Financial support from Spanish projects CICYT MAT2010-15206, JA P09-TEP-5403 and European Science Foundation (COST Action MP0805) are gratefully acknowledged. R.S thanks E. Monroy for fruitful discussion

**Figure captions**

**Fig.1** HAADF-STEM images of AlGaN section on GaN nanowires of $x_{Al}$=0.25, revealing an Al-rich shell wrapped around a Ga-rich core. The inset shows an HRTEM image at the edge of this nanowire indicated by the square taken along the [11-20] pole.

**Fig. 2** is the plot of the shell thickness ($d_{shell}$), the core ($D_{core}$) and the total diameter ($D_{tot}$) of AlGaN section as a function of distance along the nanowire growth direction. The position of GaN/Al$_x$Ga$_{1-x}$N interface is set as zero.

**Fig. 3** (a) The average Al content profile of the shell and the center nanowire measured by EDX as well as the estimated value at the core region plotted as a function of the position along the wire axis. The position of GaN/Al$_x$Ga$_{1-x}$N interface is set as zero. The lower panel shows the comparison of the predicted (b) Al content and (c) thickness with the average values from the measurement. The solid line represents the case where the predicted and the measured values are similar.

**Fig. 4** Schematic illustrations of the situation occurring during the growth of the Al$_x$Ga$_{1-x}$N section on the top of GaN nanowires: (a) the formation of Al-rich shell structures in nanowires induced by the different diffusion behavior of Al and Ga atoms and (b) the vanishing of the core-shell like structures in the Al$_x$Ga$_{1-x}$N sections caused by the shadow effect reducing the sidewall deposition.

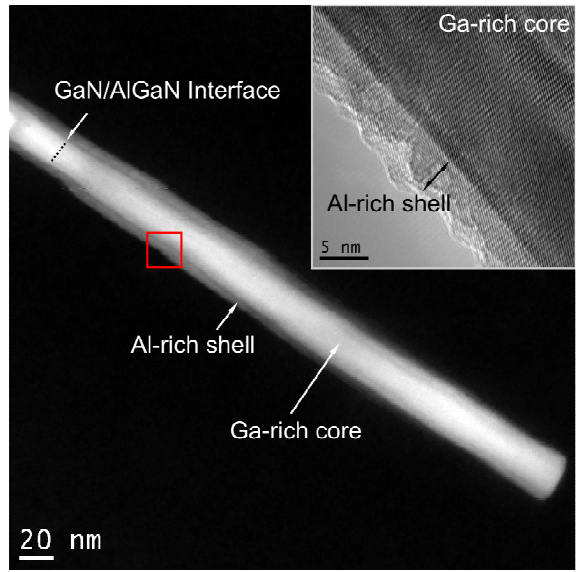

Figure 1

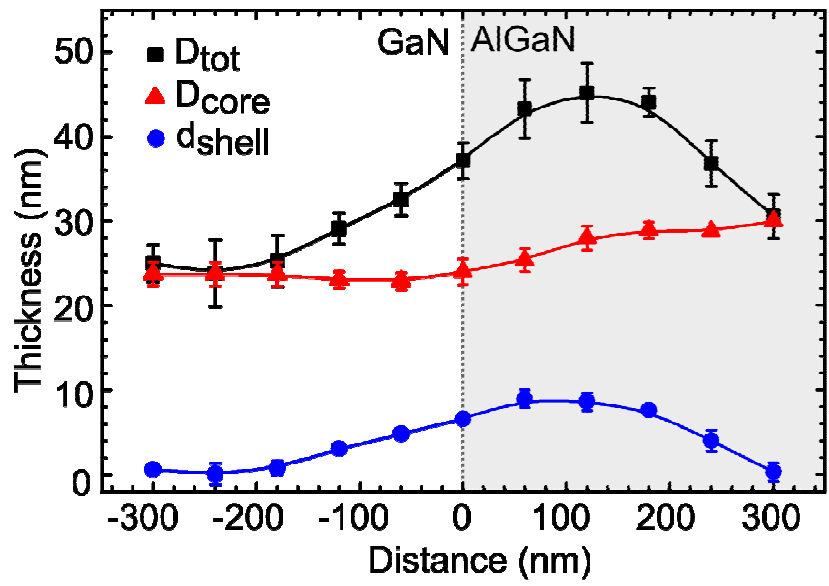

Figure 2

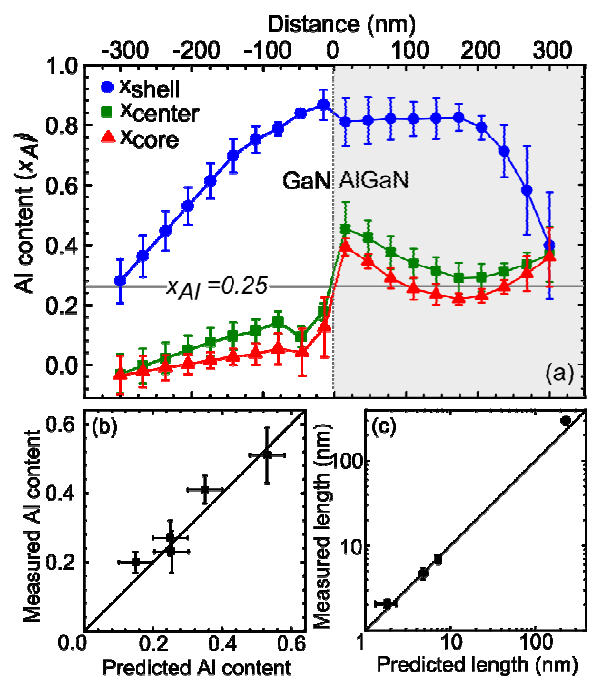

Figure 3

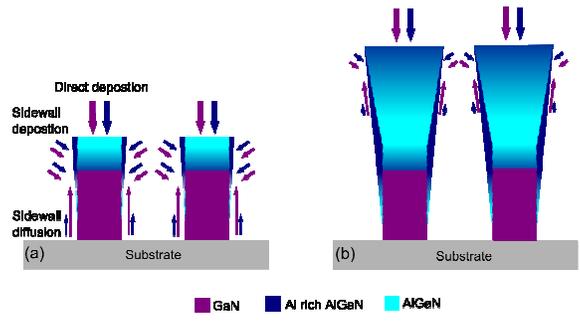

Figure 4